\documentclass[superscriptaddress,twocolumn,floatfx,amsmath,amssymb,pra]{revtex4-2}
\usepackage{hyperref}
\usepackage{xurl}
\hypersetup{breaklinks=true}
\usepackage{amsmath,amssymb}
\usepackage{graphics}
\usepackage{epsfig}
\usepackage[T1]{fontenc}
\usepackage{selinput}
\usepackage{babel}
\usepackage{ulem}
\usepackage{MnSymbol}
\usepackage{mathrsfs}
\usepackage[usenames]{color}
\usepackage{appendix}

\begin{document}

\title{Spin structure of diatomic  van der Waals molecules of alkali atoms}

\author{Jing-Lun Li}	
\affiliation{Institut f\"{u}r Quantenmaterie and Center for Integrated Quantum Science and
	Technology IQ$^{ST}$, Universit\"{a}t Ulm, 89069 Ulm, Germany}
\author{Paul S. Julienne}
\affiliation{Joint Quantum Institute, University of Maryland, and the National
Institute of Standards and Technology (NIST), College Park, MD 20742, USA}
\author{Johannes Hecker Denschlag}
\affiliation{Institut f\"{u}r Quantenmaterie and Center for Integrated Quantum Science and
	Technology IQ$^{ST}$, Universit\"{a}t Ulm, 89069 Ulm, Germany}
 \author{Jos\'{e} P. D'Incao}
\affiliation{JILA, NIST, and the Department of Physics,
University of Colorado, Boulder, CO 80309, USA}	
\affiliation{Department of Physics, University of Massachusetts Boston, Boston, MA 02125, USA}
\date{\today}

\begin{abstract}
We theoretically investigate the spin structure of weakly bound diatomic van der Waals molecules formed by two identical bosonic alkali atoms. Our studies were performed using known Born-Oppenheimer potentials while developing a reduced interaction potential model. Such reduced potential models are currently a key
for solving certain classes of few-body problems of atoms as they decrease the numerical burden on the computation. 
Although the reduced potentials are significantly shallower than actual Born-Oppenheimer potentials, they still capture the main properties of the near-threshold bound states, including their spin structure, and the scattering states over a broad range of magnetic fields.
At zero magnetic field, we find that the variation in spin structure across different alkali species originates from the interplay between electronic spin exchange and hyperfine interactions. To characterize this competition we introduce a single parameter, which is a function of the singlet and triplet scattering lengths, the atomic hyperfine splitting constant, and the molecular binding energy. 
We show that this parameter can be used to classify the spin structure of vdW molecules for each atomic species.
\end{abstract}

\maketitle
\section{Introduction}
Van der Waals (vdW) complexes are weakly bound molecular states held together by long-range dispersion interactions, in contrast to typical molecular states where the strong chemical bond originates from the overlap of the atom's electron clouds \cite{Jeziorski:1994,Reilly:2015}. These fragile molecules play a central role across a broad range of phenomena across physics, chemistry, biology, and materials science \cite{Blaney:1976, Bernstein:1995, Jones:2006, Nesbitt:2012, Hermann:2017}. Intensive investigations of vdW complexes span from spectroscopy in supersonic molecular beams \cite{Hutson:1990, Koperski:2002}, self-assembly in nanostructures \cite{Rance:2010,Gobre:2013,Carlos:2015}, molecular dynamics of biopolymers \cite{Persson:2009, Rossi:2015}, superfluidity of the Helium droplets \cite{Toennies:2004, Krzysztof:2008,Tariq:2010,Brahms:2010,Brahms:2011,Quiros:2017,Mirahmadi:2021,Mirahmadi:2021c}, to the controlled formation, manipulation and state-resolved detection of vdW molecules in 
ultracold quantum gases of alkali metal atoms
\cite{Kohler:2006,Jones:2006,Ulmanis:2012, Harter:2013,Dincao:2018,Wolf:2017, Wolf:2019, Haze:2022, Haze2023}.

An important direction of research involving vdW molecules
concerns cold, controlled chemical reactions.
Recent studies with ultracold atomic gases have revealed that the spin structure of the van der Waals bound state can play an important role in the product distribution following reactions such as three-body recombination 
\cite{Li:2022,Haze:2022,SHpaper}.
In particular,  the specific spin structure
of Rb$_2$ vdW bound states at low magnetic fields gives rise to a spin conservation propensity rule in 
three-body recombination \cite{Haze:2022, Haze2023}.
One can also expect that similar spin propensity rules 
exist for cold reactions between vdW molecules colliding with atoms and that they depend on the molecular spin structure. 
Furthermore, the spin structure of vdW molecules can vary considerably among alkali species and can ultimately influence how they can be manipulated by external fields. Therefore, determining the spin structure of vdW molecules is of fundamental importance to understand and explore various types of ultracold chemical reactions currently accessible to experiments with ultracold atoms and molecules.

Investigating the spin structure of alkali vdW diatomic molecules can be done by solving the coupled-channel Schr\"{o}dinger equation with well-known {\it ab initio} Born-Oppenheimer (BO) potentials \cite{Julienne2014,Knoop2011,Tiemann2020,Strauss2010,Berninger2013} and hyperfine interactions. However, for future implementation in reaction dynamics in three-body numerical simulations \cite{3bnum} it is also imperative to design simpler potential models to replace the actual BO potentials. In the simplified model, potentials can typically be made much shallower to alleviate the computational burden in three-body calculations \cite{Wang:2014,Chapurin:2019,Tempest:2023}. We explore this aspect in this present manuscript. Numerical simulations for three-body recombination of ultracold atoms, which use the actual BO potentials, have been conducted in Refs. \cite{Secker:2021, Li:2022,kraats:2024}. In such studies, additional high momentum truncations were implemented in momentum-space \cite{Secker:2021, Li:2022,kraats:2024}, which has a similar effect to reducing the depth of potential in coordinate space. Nevertheless, developing potential models with reduced depth is more suitable for implementation in coordinate-space approaches.

In this work, we investigate the vdW molecules of two identical bosonic 
alkali atoms for $^7$Li, $^{23}$Na, $^{39}$K, $^{41}$K, $^{85}$Rb, $^{87}$Rb 
and $^{133}$Cs and characterize the spin structure
of these molecules states. For these molecules, their spin structure emerges from a combination of various spin-dependent interactions, including electronic spin exchange and hyperfine interactions. We find that, at zero magnetic field, the molecular spin structure is 
determined by the ratio of the energies characterizing the electronic spin exchange and hyperfine interactions. This ratio varies across different species and can be qualitatively described 
by a single parameter, determined by the singlet 
and triplet scattering lengths, the atomic hyperfine splitting 
constant, and the molecular binding energy. We construct reduced potential models for such diatomic vdW systems that are shallower but largely preserve the major properties of the actual interactions. We aim to design for each species a potential model that is generally good for representing the binding energy and spin structure of weakly bound vdW molecules. We show that the reduced potentials also describe well the low-energy scattering property of two atoms. In fact, we find that the latter criterion can be used as a practical way to construct the reduced potential so that the former criterion is automatically satisfied. We note that our reduced potential model simultaneously captures the multichannel low-energy spectrum and scattering properties by incorporating the realistic atomic spin structure. The rest of this paper is organized as follows: in Section II we introduce our theoretical framework, characterize the range of binding energies where the molecular states can be considered as vdW molecules and detail our procedure to construct reduced model potentials. In Section III we compare the results of reduced and realistic potentials. We characterize the spin mixing of vdW molecules in Section IV and summarize our main findings and conclusion in Section V. 

\section{Theoretical Framework} \label{sec:theo}
For two alkali atoms in an external magnetic field
the Hamiltonian reads
\begin{equation}
\hat{H}=\hat{T}(r)+\hat{V}(r)+\hat{H}_{\rm hf}+
\hat{H}_{\rm Z}, \label{Ham}
\end{equation}
where $\hat{T}$ and $\hat{V}$ denote the kinetic and potential energy operators, respectively, and $r$ is the interatomic distance. The third term above represents the sum of the two (identical) atomic hyperfine interactions, $\hat{H}_{\rm hf}=A_{\rm hf}(\vec{s}_{a}\cdot \vec{i}_{a})/\hbar^{2}+A_{\rm hf}(\vec{s}_{b}\cdot \vec{i}_{b})/\hbar^2$, where $A_{\rm hf}$ is the hyperfine constant, and $\vec{s}_{i}$ and $\vec{i}_{i}$ the electronic and nuclear spins, respectively, of atom $i$ ($i=a,b$). The fourth term describes the Zeeman Hamiltonian $\hat{H}_{\rm Z}=(\gamma_{\rm e} \vec{s}_a+\gamma_{\rm e} \vec{s}_b-\gamma_{\rm n} \vec{i}_a-\gamma_{\rm n} \vec{i}_b)\cdot \vec{B}$ in the presence of a homogeneous magnetic field in the $\hat{z}$, $\vec{B}=B\hat{z}$, with $\gamma_{\rm e}$ and $\gamma_{\rm n}$ being the electronic and nuclear gyromagnetic factors, respectively. The physical parameters for the alkali atoms discussed above can be found in Ref. \cite{Arimondo:1977}

\begin{figure}[htbp]
 \centering
  \resizebox{0.48\textwidth}{!}{\includegraphics{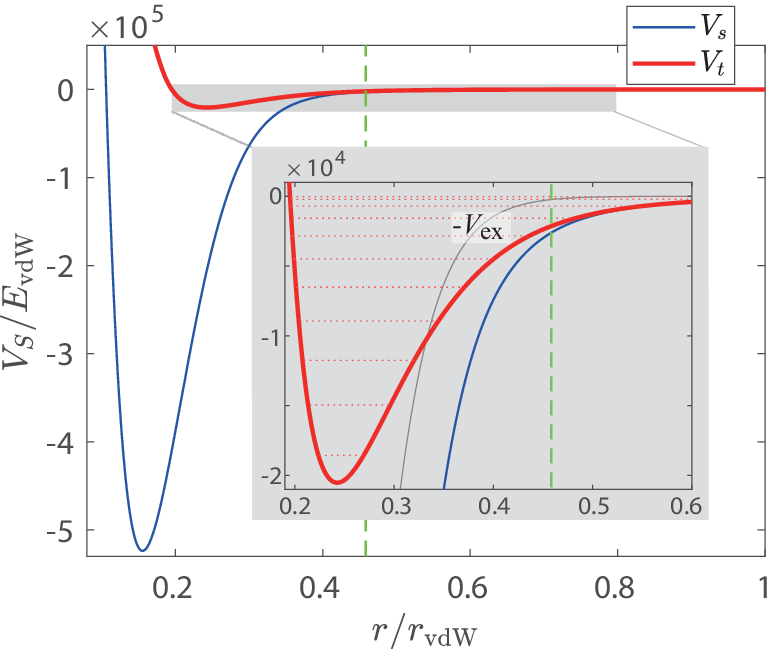} }
 \caption{\label{fig:pot} Singlet (thin blue) and triplet (thick red) BO potentials of $^{7}$Li$_2$ are displayed in vdW units. The vertical dashed line represents a critical internuclear distance, $r_{\rm ex}$, defined in the text for classifying the vdW molecules. The inset provides a zoomed-in view of the shaded region, showing 11 triplet s-wave bound levels marked by dotted lines. The thin grey line displays the electronic spin-exchange interaction $-V_{\rm ex}$. }
\end{figure}

Here, we note several properties of the molecules under consideration that simplifies the Hamiltonian to the form of Eq. (\ref{Ham}). Firstly, the considered molecules consist of two ground-state alkali-metal atoms with no electronic orbital angular momentum, excluding interactions such as electron spin-orbital coupling and nuclear spin-electron orbital couplings. Secondly, the magnetic dipole-dipole interaction of electron spins and the second-order spin-orbit interaction typically have only a small effect \cite{chin:2010}. However, including these terms in the Hamiltonian can be crucial under special conditions, such as in the case of Cs$_2$ discussed later. Nevertheless, in this work, we generally exclude such interactions. Finally, the molecules are weakly bound, making the relevant physics predominantly long-range. Consequently, the hyperfine interaction can be reasonably taken as the Fermi contact form with a constant $A_{\rm hf}$. A $r$-dependent modification that captures the short-range cross-atom hyperfine interaction and other terms (as introduced in Ref. \cite{Strauss2010} ) is not necessary. At a large internuclear distance the spin-nuclear rotation coupling strength also becomes weak \cite{Veseth:1971,zhu:2019}, resulting in a negligible spin-nuclear rotation interaction. As a result, we can write the total wave function $\Psi(\vec{r})$ as
\begin{align}
    \Psi(\vec{r})=\sum_{\alpha}\frac{\psi_\alpha(r)}{r}Y_{lm_l}(\theta,\phi)|\alpha\rangle,\label{wf}
\end{align}
where $Y_{l m_l}$ is the spherical harmonics for the orbital angular moment $l$ 
and azimuthal projection $m_l$, $|\alpha\rangle$ is the eigenstate of 
$\hat{H}_{\rm hf}+\hat{H}_{\rm Z}$ with eigenvalue $E_\alpha$, and $\psi_\alpha(r)$ is the radial wavefunction. At zero magnetic field,  $|\alpha\rangle=|f_am_{f_a}\rangle|f_bm_{f_b}\rangle$ is the direct product of the eigenstates $|fm_f\rangle$
of the atomic hyperfine spins ($\vec{f}=\vec{s}+\vec{i}$) where $f$ is the hyperfine spin and $m_f$ its azimuthal projection. Note that here we will use $|f_am_{f_a}\rangle|f_bm_{f_b}\rangle$ to label $|\alpha\rangle$ states even when a magnetic field is non-zero.
At each magnetic field $B$, we then solve the coupled channel Schr\"{o}dinger equation
given by
\begin{align}
    \big[-\frac{\hbar^2}{m}\frac{d^2}{dr^2}
    &+\frac{l(l+1)}{m r^2}\hbar^2+E_\alpha\big]\psi_{\alpha}(r)
    \nonumber\\
    &+\sum_{\alpha'}{\langle\alpha|\hat{V}(r)|\alpha'\rangle}\psi_{\alpha'}(r)=E\psi_{\alpha}(r),\label{Schro}
\end{align}
where $m$ is the atomic mass.
For alkali atoms, the interatomic interactions depend on the electronic spins and we represent the potential energy operator as
\begin{equation}
\hat{V}(r)=
\sum_{SM_S}|SM_S\rangle V_S(r)\langle SM_S|,\label{VS}
\end{equation}
where $S$ is the total electronic spin, $|s_a-s_b|\le S \le {s}_a+{s}_b$, and $-S\le M_S\le S$ is azimuthal projection.
For alkali atoms the electronic spins allow for the singlet ($s$), $S=0$, and triplet ($t$), $S=1$, BO interactions. We use $s$ and $t$ as indices for $S=0$ and 1, respectively, in the following. For all species considered in this study, the actual BO potentials of two identical atoms are considerably deep, typically supporting $\sim 10$ to $\sim 100$ $s$-wave ($l=0$) bound states  \cite{Julienne2014,Knoop2011,Tiemann2020,Strauss2010,Berninger2013}, with a total number of rovibrational states ranging from hundreds up to a few thousands. Our calculations show that the least number of the $s$-wave bound states is 11, arising from the triplet potential of $^{7}$Li$_2$ while the highest number is 156 from the singlet potential of $^{133}$Cs$_2$. Figure \ref{fig:pot} shows the BO potentials (taken from Ref. \cite{Julienne2014}) and the $s$-wave bound levels of $^7$Li$_2$ to illustrate their general form. The two potentials deviate at short range due to electronic spin-exchange interactions but become nearly identical beyond a critical internuclear distance, $r_{\rm ex}$, which will be defined in the following. The dominant contribution to the long-range part of the BO potentials is given by the $-C_6/r^6$ dispersion interaction. Here, $C_6$ is the vdW dispersion coefficient, from which we define the characteristic length $r_{\rm vdW}=(mC_6/\hbar^2)^{1/4}/2$ and energy $E_{\rm vdW}=\hbar^2/mr_{\rm vdW}^2$ scales of the vdW interaction. These energy and length scales define the regime where quantum threshold modification of scattering occurs \cite{chin:2010}. Furthermore, the typical size of all bound states in a vdW potential is on the order of or less than $r_{\rm vdW}$, except for the last s-wave bound state when the scattering length is large compared to $r_{\rm vdW}$. We will use vdW units throughout the following discussion unless otherwise specified.

\subsection{Criteria for vdW molecular states}

  \begin{table*}[t]
  \tabcolsep=4pt
\small
\renewcommand\arraystretch{1.0}
\caption{\label{tab:vdWM}  Parameters for classifying vdW molecules composed of two identical bosonic alkali atoms}
\begin{tabular}{ccccccccc}
 \hline
 \hline
 Molecule&$r_{\rm vdW}/a_0$&$r_{\rm ex}/r_{\rm vdW}$&$r_{\rm ex}/a_0$&$E_{\rm vdW}/h$ [MHz]&$E^{\rm max}_s/E_{\rm vdW}$&$E^{\rm max}_t/E_{\rm vdW}$ &$E^{\rm max}_s/h$ [THz]&$E^{\rm max}_t/h$ [THz]\\ 
 \hline
 $^{7}$Li$_2$&32.49&0.4583&14.89&487.48&2617&2141&1.28&1.04\\
 $^{23}$Na$_2$&44.96&0.3483&15.66&77.67&14105&11538&1.10&0.90 \\
 $^{39}$K$_2$&64.61&0.2703&17.46&22.19&67097&54911&1.49&1.22\\
 $^{41}$K$_2$&65.42&0.2669&17.46&20.59&72405&59242&1.49&1.22\\
 $^{85}$Rb$_2$&82.16&0.2189&17.99&6.30&245495&200822&1.54&1.26 \\
 $^{87}$Rb$_2$&82.64&0.2177&17.99&6.08&253623&207567&1.54&1.26\\
 $^{133}$Cs$_2$&101.07&0.1863&18.82&2.66&661003&540800&1.76&1.44\\
 \hline
 \hline
 \end{tabular}
 \end{table*}
As we mentioned above, vdW molecules are weakly bound molecular states held together by long-range dispersion interactions. For such molecular states, the electronic cloud for each of the atoms have little to none overlap, making vdW molecules fundamentally different to typical molecular states where the strong chemical bound originates from the overlap of the electronic clouds \cite{Jeziorski:1994,Reilly:2015}. 
 In order to characterize the range of binding energies for vdW molecules, 
 we can estimate the dominant role of the vdW interaction at different interatomic distances, thus allowing us to estimate the size (and energy) of the molecular states where vdW interactions prevails over the electronic exchange. 
The dominance of the vdW interaction can be roughly estimated via the ratio
\begin{align}
    \sigma_{\rm ex}(r) \equiv \frac{V_{\rm vdW}(r)}{V_{\rm ex}(r)}= \frac{V_s(r)+V_t(r)}{V_s(r)-V_t(r)}
\end{align}
by assuming $V_{s/t} \approx V_{\rm vdW}\mp V_{\rm ex}$, where $V_{\rm ex}$ and $V_{\rm vdW}$ denote the electronic spin-exchange interaction and the vdW interaction respectively. We define the critical internuclear distance $r_{\rm ex}$ by setting  $\sigma_{\rm ex} (r_{\rm ex})=10$, ensuring that $V_{\rm vdW}\gg V_{\rm ex}$ when $r\geq r_{\rm ex}$. We note that $r_{\rm ex}$ has a similar physical interpretation as the LeRoy radius from LeRoy-Bernstein’s theory \cite{Leroy:1974}, i.e., it indicates the internuclear distance at which the electron cloud overlap is negligible. The dominance of vdW interaction at $r\geq r_{\rm ex}$ is demonstrated in Fig. \ref{fig:pot}, using $^7$Li$_2$ as an example (see Table~\ref{tab:vdWM} for the corresponding values of $r_{\rm ex}$.) Accordingly, a molecule can be classified as a vdW molecule if its size exceeds $r_{\rm ex}$ or, equivalently, its binding energy is smaller than the critical value defined as $E_S^{\rm max}\equiv |V_S(r_{\rm ex})|$. The critical values for the molecular binding energies are listed in Table \ref{tab:vdWM}. We find that the $E_s^{\rm max}$ and $E_t^{\rm max}$ are typically around 1 THz, corresponding to a $r_{\rm ex}$ of approximately 15-19 $a_0$. Nevertheless, as we can see from Table~\ref{tab:vdWM}, $^7$Li is the atomic species with the most restrictive range of binding energies in vdW units, with a critical value of a few thousand $E_{\rm vdW}$. Therefore, in this work, we will only consider molecular states with a binding range of up to a few thousand $E_{\rm vdW}$ to ensure that the vdW interaction dominates molecular bond formation for all species explored.

\subsection{Constructing reduced potential models}

This section aims at introducing a reduced interaction model to replace $V_S$ in Eq.~(\ref{VS}), containing a much smaller number of molecular states and denoted by $V_S^*$. Both reduced singlet ${V}_s^*$ and triplet ${V}^*_t$ potentials should be sufficiently shallower than their originals and reproduce well the low-energy bound and scattering properties of the system. Here, we define the reduced potentials as
\begin{equation}
{V}^*_{S}(r)=V_{S}(r)+\frac{C_6 \lambda_{S}^6}{r^{12}},\label{redpot}
\end{equation}
where $V_S$ is the actual BO potential and $\lambda_{S}$ is a free (length) parameter used to tune the introduced short-range $1/r^{12}$ repulsion. This effectively reduces the number of bound states. 

  
  \begin{table*}[htbp]
  \tabcolsep=4pt
\small
\renewcommand\arraystretch{1.0}
\caption{\label{tab:rp} Parameters for the reduced potentials, $n_{s/t}$, $\lambda_{s/t}$, $\lambda_{s/t}^*$, and $c_{\rm hf}$ (see text), along with the relevant parameters characterizing the corresponding atomic species, $a_{s/t}$ and $A_{\rm hf}$. See Table~\ref{tab:vdWM} for the corresponding values of $r_{\rm vdW}$. Note that here we list $A_{\rm hf}$ with only two decimal places, while in our calculation the precise value from Ref. \cite{Arimondo:1977} is employed.}
\begin{tabular}{cccccccccccc}
 \hline
 \hline
 Molecule& $a_{s}$/$a_0$&$n_s$&$\lambda_s/r_{\rm{vdW}}$&$\lambda^*_s/r_{\rm{vdW}}$& $a_{t}$/$a_0$&$n_t$&$\lambda_t/r_{\rm{vdW}}$&$\lambda_t^*/r_{\rm{vdW}}$&$A_{\rm hf}/h$ [MHz]&$c_{\rm{hf}}$ \\ 
 \hline
 $^{7}$Li$_2$&34.34&6&0.3792958&$\lambda_s$&-26.85&5&0.3181707&$\lambda_t$&200.88&1\\
 $^{23}$Na$_2$&18.83&6&0.3395979&$\lambda_s$&64.31&6&0.3235117&$\lambda_t$&442.91&1 \\
 $^{39}$K$_2$&138.89&6&0.3272250&0.3269770&-33.28&5&0.3354390&0.3354690&115.43&1 \\
 $^{41}$K$_2$&85.43&6&0.3226359&0.3225499&60.29&5&0.3437314&0.3437234&63.50&1 \\
 $^{85}$Rb$_2$&2654.70&6&0.3242030&$\lambda_s$&-389.63&5&0.3257500&0.3258900&505.96&0.9477 \\
 $^{87}$Rb$_2$&90.20&6&0.3131870&0.3133710&98.91&6&0.3138290&0.3140510&1708.67&0.9026 \\
 $^{133}$Cs$_2$&286.52&7&0.2946725&0.2896405&2857.28&7&0.2974180&0.2993080&1149.08&0.8710\\
 \hline
 \hline
 \end{tabular}
 \end{table*}

 \begin{figure}[t]
 \centering
  \resizebox{0.48\textwidth}{!}{\includegraphics{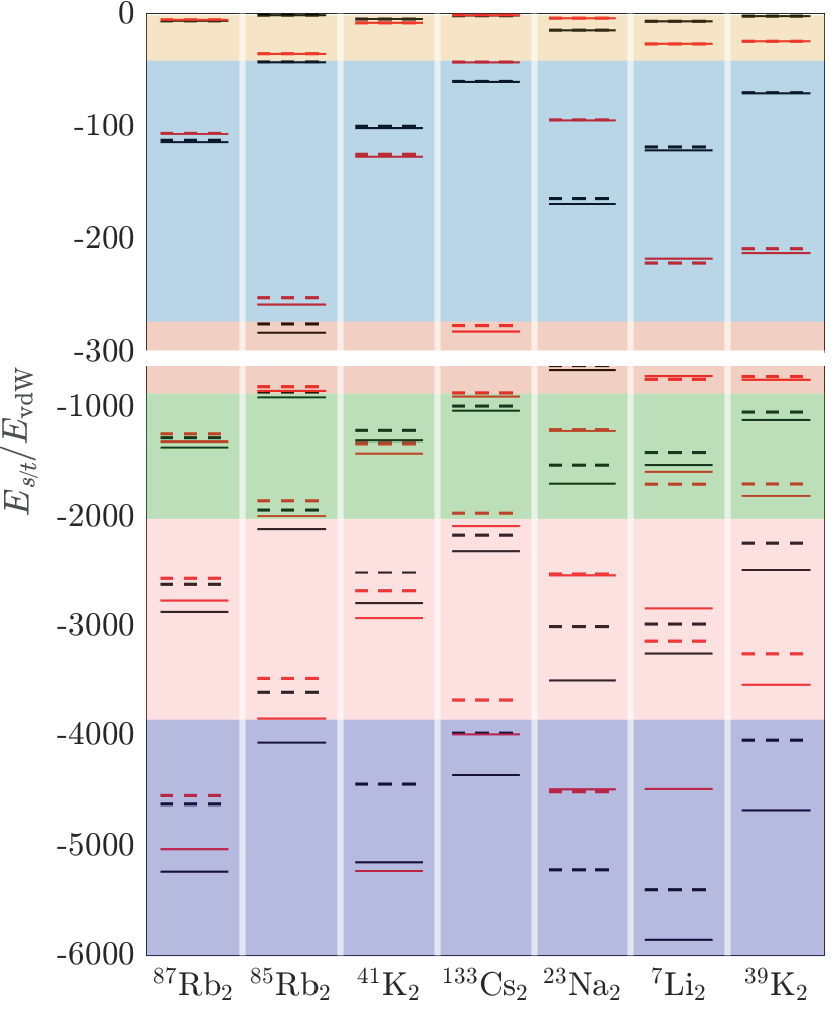} }
 \caption{\label{fig:EST} Singlet $E_s$ (black) and triplet $E_t$ (red) molecular energies from the BO potentials (solid lines) and from the corresponding reduced (dashed lines) potentials (see Table \ref{tab:rp}) for bosonic alkali dimers. The alkali species are arranged in order of increasing effective singlet-triplet level splitting $u_{st}$. The colored regions represent different `energy bins'. Both $u_{st}$ and `energy bin' are defined later in Section \ref{SpinMix}A.  The upper panel highlights the most weakly bound molecular states displaying a good agreement between the energies from the BO and reduced potentials. This agreement deteriorates for the more deeply bound molecular states displayed in the lower panel.}
\end{figure}

Our reduced potentials are determined by tuning $\lambda_S$ to the values of $\lambda_s$ and $\lambda_t$ that reproduce the values of the singlet $a_s$ and triplet $a_t$ scattering lengths, respectively, obtained from the corresponding BO potentials \cite{Julienne2014,Knoop2011,Tiemann2020,Strauss2010,Berninger2013}. [Values of $a_s$ and $a_t$ were calculated by simply setting $E_\alpha=0$ and $\hat{V}={V}_{s/t}$ or ${V}^*_{s/t}$ in Eq.~(\ref{Schro}).] 
Evidently, there are multiple choices of $\lambda_s$ and $\lambda_t$ that reproduce the values of $a_s$ and $a_t$, each corresponding to a different number of $s$-wave bound states that the reduced potentials can support. For our present study, we choose the number of $s$-wave singlet states to be $n_s=6$, except for $^{133}$Cs, where we choose $n_s=7$ (see later discussion). For the reduced triplet interaction, this number is chosen as $n_t=n_s$ for cases where $a_s<a_t$, or $n_t=n_s-1$ for cases where $a_s>a_t$. This ensures that the singlet potential is deeper than the triplet potential in the reduced potential model, consistent with the actual BO potentials. According to our analysis in Section \ref{sec:theo}A, all the levels included in the reduced potentials are vdW molecules except for the last two singlet and the last triplet $^{7}$Li$_2$ levels. Including deeper bound state levels, particularly for $^7$Li, is beyond the scope of the present work as the vdW interaction may no longer be dominant in a molecular bond formation.
The values of $\lambda_s$ and $\lambda_t$, and corresponding scattering lengths are listed in Table \ref{tab:rp}. 

Figure \ref{fig:EST} (upper panel) compares the energies of the first few 
singlet and triplet bound state levels, $E_s$ and $E_t$, respectively, 
of the reduced potentials (dashed 
lines) to those obtained from the original BO potentials (solid lines). 
The figure shows that by matching the singlet and triplet scattering lengths 
one obtains a very good agreement between the energy levels to the results from BO potentials. 
Evidently, such agreement deteriorates for more deeply bound levels 
as shown in the lower panel of Fig. \ref{fig:EST}. Such deterioration 
originates from the fact that the reduced potentials are much shallower than 
the BO potentials. This can be further improved by using deeper reduced 
models containing a larger number of molecular states.
Figure \ref{fig:EST} also clearly shows that each colored block contains 
exactly one bound level for each potential (note that the levels in the energy range of [-600, -300] $E_{\rm vdW}$ are not shown due to a break in the y-axis). As we shall discuss in more detail later in 
Section \ref{SpinMix}, these colored blocks represent a universal structure 
of the dimer spectrum of vdW potentials, conventionally referred to as the 
vdW energy bins \cite{chin:2010, Gao:2000}.

Throughout this work, we use the log-derivative algorithm \cite{Manolopoulos:1986} to solve the Schr\"{o}dinger equation (\ref{Schro}) for calculating the low-energy scattering quantities. We switch to a mapped grid Hamiltonian method \cite{Willner:2004} when bound level energies or the scattering and bound state wavefunctions are required.
\section{Performance of the reduced potentials}

Although we find that the reduced singlet and triplet potentials reproduce well the molecular energies, for applications in ultracold atoms the actual performance of the reduced potential model needs to be reevaluated in the presence of hyperfine and Zeeman
interactions, $\hat{H}_{\rm hf}$ and $\hat{H}_{\rm Z}$ in Eq.~(\ref{Ham}).
In fact, we find that the reduced potentials need to be fine-tuned to precisely describe the low energy scattering properties of the system at finite $B$-fields for some species.
In this section, we describe such adjustments and evaluate the performance 
of the reduced potentials in comparison to the BO potentials by analyzing the 
corresponding scattering and bound properties as well as the spin structure of both scattering and molecular states.

\subsection{Scattering properties}

For our present study, we consider that the colliding atoms are prepared in 
the hyperfine spin-stretched state, $|f$=$f_*$,$m_f$=$\text -f_*\rangle$, of the lowest 
hyperfine manifold. Concretely, $f_*=3$ for $^{133}$Cs, $f_*=2$ for 
$^{85}$Rb and $f_*=1$ for all other atomic species. 
We focus on the low magnetic field range of [-200, 200] G, which is the 
typical accessible experimental regime of Refs. \cite{Wolf:2017,Wolf:2019, 
Haze:2022,Haze2023} and for many other experiments. We note that considering the $|f=f_*,m_f=\text-
f_*\rangle$ state at a negative magnetic field is equivalent to studying the 
physics of the $|f=f_*,m_f=+f_*\rangle$ state at $|B|$. Since the reduced models obtained in the previous section do not include the effects of hyperfine and Zeeman interactions further adjustments might be necessary to better fit the $B$-field dependency of the relevant physical observables. Here, we choose to use the low energy scattering properties of the system, parameterized by the $s$-wave scattering length $a(B)$ and effective range $r_e (B)$, to establish such adjustments. At each $B$-field, we solve the radial coupled channel Schr\"{o}dinger equation (\ref{Schro}) while varying the collision energy, $E$. By determining the energy-dependent phase shift $\delta(k)$, where $k=\sqrt{mE}/\hbar$, we extract $a (B)$ and $r_e (B)$ from the effective range expansion
\begin{equation}
k\cot \delta(k)=-\frac{1}{a}+\frac{1}{2}r_e k^2+{\cal O}(k^4).
\end{equation}

In order to precisely reproduce the values of $a(B)$ and $r_e(B)$ obtained from the BO potentials, we allow for small variations the values of $\lambda_s \rightarrow \lambda_s^*$ and $\lambda_t \rightarrow \lambda_t^*$ controlling the reduced potentials [Eq.~(\ref{redpot})]. 
In addition to that, for heavier atomic species where the hyperfine splitting constant is typically large, we will also allow for slight changes the value of the hyperfine constant $A_{\rm hf}$ in Eq.~(\ref{Ham}). This new hyperfine constant is given by $A_{\rm hf}^*=c_{\rm hf}A_{\rm hf}$, where $c_{\rm hf}$ should be kept close to 1 to preserve the main spin structure of the system. 

  \begin{figure}[t]
 \centering
  \resizebox{0.48\textwidth}{!}{\includegraphics{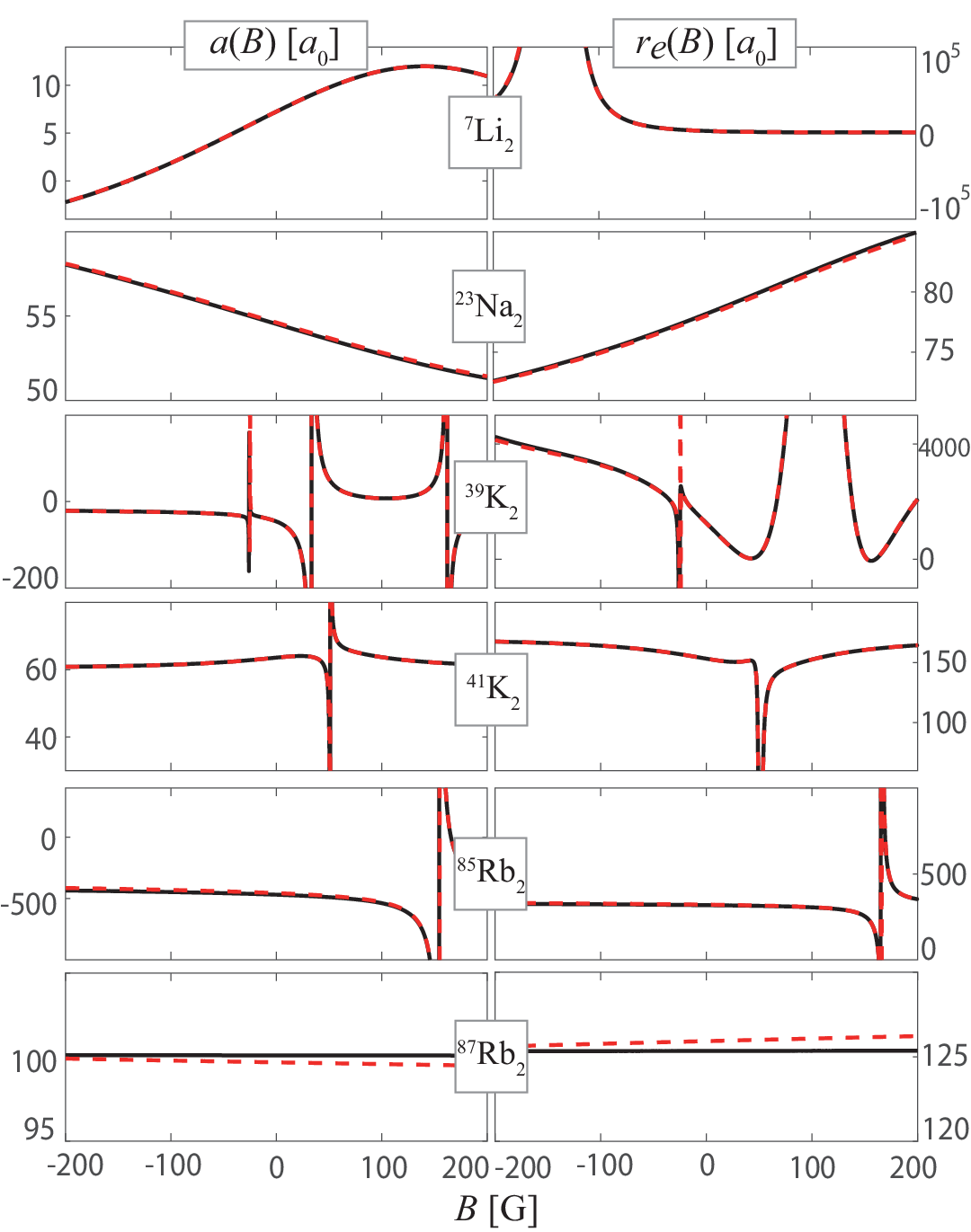} }
 \caption{\label{fig:ARE} Scattering length $a(B)$ and effective range $r_e(B)$ of two identical bosonic alkali atoms from original (solid lines) and reduced potentials (dashed lines). In our case, the discrepancy between the results from the reduced and BO potentials are similar across species, typically on the order of 0.1$\%$ to 1$\%$. It should be noted that we use $|f^*,m_f=\text-f^*\rangle$ as the reference spin state for the whole considered magnetic field range, which means that a negative magnetic field value represents the $|f^*,m_f=f^*\rangle$ state at $|B|$.}
\end{figure}

 \begin{figure}[t]
 \centering
  \resizebox{0.47\textwidth}{!}{\includegraphics{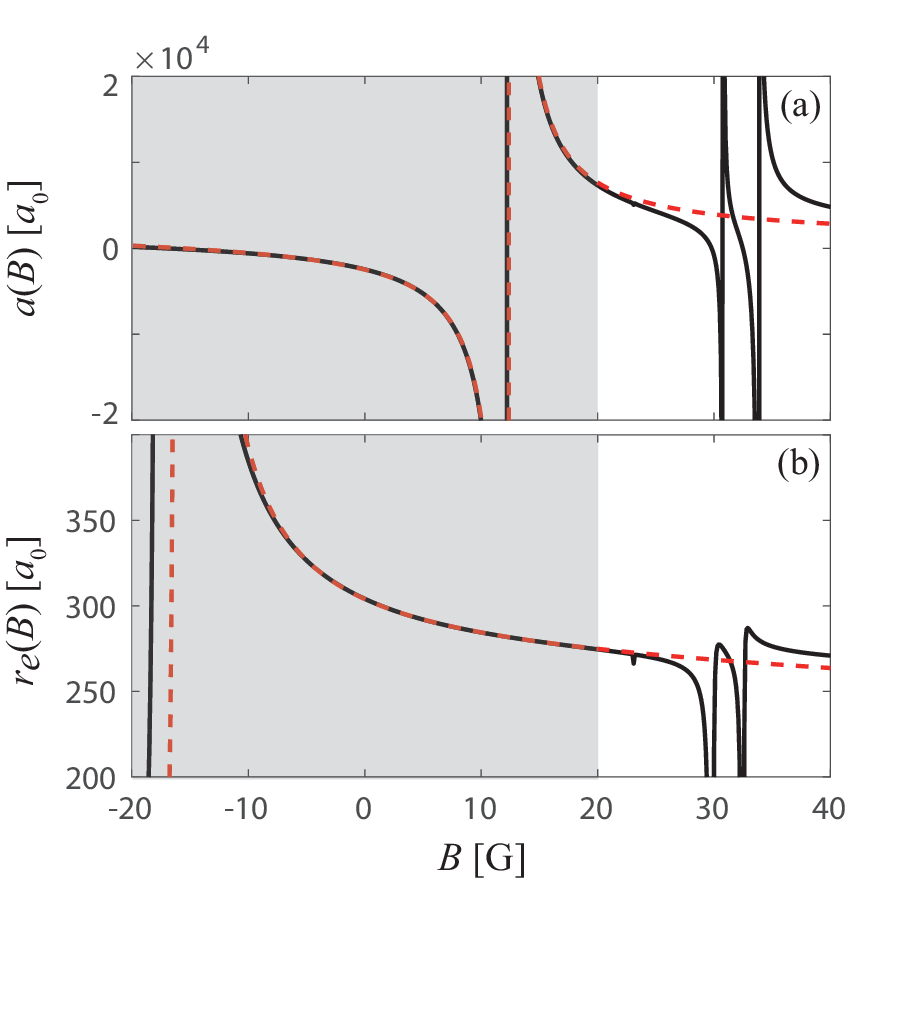} }
 \caption{\label{fig:133CS} The $s$-wave two-body scattering length (a) and effective range (b) for $^{133}$Cs$_2$ from the original potentials (solid line) and reduced potentials (dashed line). For the calculation with original potentials we include the dipole-dipole interaction and take both the $s$ and $d$ partial wave into account, while for that with reduced potentials, we neglect the dipole-dipole interaction. The shaded area indicates the considered magnetic field regime for fine-tuning $\lambda_s$ and $\lambda_t$. It should also be noted that we use $|f^*,m_f^*=\text-f^*\rangle$ as the reference spin state for the whole considered magnetic field range, which means that a negative magnetic field value represents the $|f^*,m_f=f^*\rangle$ state at $|B|$.}
\end{figure}

The resulting performance of reduced potentials in describing the magnetic field-dependent scattering properties is demonstrated in Figs. \ref{fig:ARE} and \ref{fig:133CS} for the adjusted model parameters $\lambda_{s/t}^*$ and $c_{\rm hf}$ listed in Table.~\ref{tab:rp}. For all species, the $a(B)$ and $r_e(B)$ are very well reproduced by the reduced potentials in [-200, 200] G, as compared to the result from the original BO potentials. Remarkably, the features of the Feshbach resonances and the zero crossing of $a(B)$ [or equivalently, the divergence of $r_e(B)$] in $^{7}$Li$_2$, $^{39}$K$_2$, $^{41}$K$_2$ and $^{85}$Rb$_2$ are correctly captured (see Fig.~\ref{fig:ARE}). These features are important ingredients for investigating three-body problems regarding, for instance, the Efimov effect \cite{Chapurin:2019,Xie:2020,Secker:2021,Wild:2012} and the control of three-body reaction via a magnetic field \cite{FRpaper}. To achieve such a good agreement, fine-tuning in $\lambda_s$ and $\lambda_t$ is either not necessary (as for $^{7}$Li$_2$ and $^{23}$Na$_2$, and the singlet potential of $^{85}$Rb$_2$), or required to less than $0.1\%$. The hyperfine constant is reduced by about $5\%$ and $10\%$ for $^{85}$Rb$_2$ and $^{87}$Rb$_2$, respectively. As a result, the number of $s$-wave bound states, $n_s$ and $n_t$, is not affected while shifts in the molecular energies (compared to those in Fig.~\ref{fig:EST}) and in $a_s$ and $a_t$ (if any) are typically within a few percent.

The adjustment of parameters for the heaviest atomic species $^{133}$Cs (see Fig.~\ref{fig:133CS}) is slightly different. The low $B$-field scattering properties of $^{133}$Cs$_2$ in $|f_*m_f^*\rangle|f_*m_f^*\rangle$ state are strongly affected by multiple Feshbach resonances. We emphasize that the first resonance at $B=12$ G originates from an $s$-wave molecular state. In addition, there are resonances at $B=31$ G and 33 G caused by a $d$-wave molecular level crossing the $s$-wave threshold. The latter resonances occur due to the couplings from the magnetic dipole-dipole interaction \cite{Berninger2013}. As a result, a thorough theoretical model for $^{133}$Cs$_2$ would need to incorporate the magnetic dipole-dipole interaction into Hamiltonian (\ref{Ham}) in order to allow for the coupling between $s$ and $d$ partial waves. This would add a significant degree of complexity in both two- and three-body numerical simulations \cite{wang2011PRLa,wang2011PRLb}.
Since our goal is to develop a simple reduced potential model for further uses in three-body calculations, we include the magnetic dipole-dipole interaction for $^{133}$Cs only when we calculate $a(B)$ and $r_e(B)$ from the BO potentials and adjust the reduced potentials to fit these results. 
As a result, we only accurately reproduce the $s$-wave resonance at $B=12$ G. This resonance is caused by a $v=-7$ closed-channel bound level crossing the open-channel threshold. It is crucial to include this level in the reduced potentials to properly capture the resonance feature. Therefore, we use a larger number of bound states, i.e., $n_s=7$, for $^{133}$Cs$_2$ than other species. Since our reduced potential model can not describe the $d$-wave resonances at 31 G and 33 G, the fitting should also avoid the corresponding magnetic field regime. We fine-tune the short-range potential parameters, $\lambda_s$ and $\lambda_t$, to fit the reference values of $a(B)$ and $r_e(B)$ in the range of [-20, 20] G for $^{133}$Cs$_2$, with $c_{\rm hf}$ also being adjusted (see Table \ref{tab:rp}). Figure \ref{fig:133CS} shows the good performance of our obtained reduced potential in reproducing the physical $a(B)$ and $r_e(B)$ in this magnetic field regime. However, in the range of [20, 40] G, the result of the reduced potentials deviates from that of the original BO potentials. We note that the fine-tuning of $\lambda_s$ and $\lambda_t$ for $^{133}$Cs is relatively large ($1\% \sim 2\%$) as compared to other species, leading to $n_s\rightarrow n_s-1$ while $n_t$ is still unchanged. We attribute this to the omission of dipole-dipole interaction. The reduction of hyperfine constant by $c_{\rm hf}$ is about $13 \%$.

\subsection{Bound state properties}
 \begin{figure}[t]
 \centering
  \resizebox{0.48\textwidth}{!}{\includegraphics{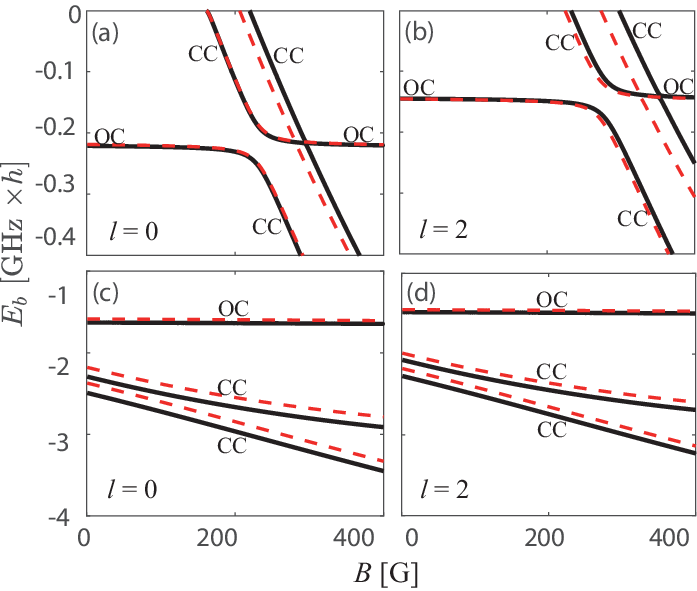} }
 \caption{\label{fig:85Rb2} The energy of $s$-wave, (a) and (c), and $d$-wave, (b) and (d), molecular levels of $^{85}$Rb$_2$ obtained from the original potentials (solid lines) and reduced potentials (dashed lines). We use `OC' or `CC' to indicate that the molecular level belongs to the incoming open channel or a closed channel, respectively. Note that such a classification does not apply in the vicinity of an avoid crossing. }
\end{figure}

Weakly bound vdW molecules are fragile and highly susceptible to external perturbations. Consequently, an external field, such as a magnetic field, can significantly alter their loose bonds, offering a valuable opportunity to control the properties of these molecules. This subsection examines the response of weakly bound vdW molecular levels to a magnetic field and assesses the performance of the reduced potentials. The success of the reduced potentials in accurately reproducing $a(B)$ and $r_e(B)$ suggests that these potentials are also capable of reproducing the bound levels. Here, we demonstrate this in more detail by using the results obtained for $^{85}$Rb$_2$ as an example. In Fig.~\ref{fig:85Rb2} we compare a group of $l=0$ and 2 bound level energies obtained from the reduced and BO  potentials. For the most weakly bound states [Fig.~\ref{fig:85Rb2}(a) and \ref{fig:85Rb2}(b)], the corresponding energies for both partial waves are in general well reproduced by our reduced potential in a wide range of $B$-field [0, 400] G, i.e, even beyond the regime considered in the fitting process discussed in the previous section. As expected, deviations become more perceptible for deeper molecular levels, as shown in Fig.~\ref{fig:85Rb2}(c) and \ref{fig:85Rb2}(d). In Fig. \ref{fig:85Rb2}, the molecular levels that are independent of the magnetic field correspond to the incoming channel levels. In contrast, the other molecular levels, which are strongly dependent on the magnetic field, are associated with closed channels. The closed channels have a magnetic moment difference relative to the incoming channel. The magnetic moments of these molecular states, reflected in the slopes of the energy curves, are largely conserved, except when strong couplings between molecular levels occur by chance, such as during an avoided crossing. We shall discuss more details of the avoided crossing in the next subsection. A closed channel molecular state is more deeply bound (by the corresponding threshold separation) than an incoming channel molecular level at the same energy. Consistently, deviations are also more perceptible for the closed channel molecular states in Fig. \ref{fig:85Rb2}. 
We note that our reduced potentials can typically support bound states in a partial wave of up to $l=20$. We have also checked the energies of states of $l>2$ partial waves and found that the performance of the reduced potential is generally good for small $l$.
We observe similar results for the other atomic species studied here. 
\subsection{Spin structure}
 \begin{figure}[t]
 \centering
  \resizebox{0.4\textwidth}{!}{\includegraphics{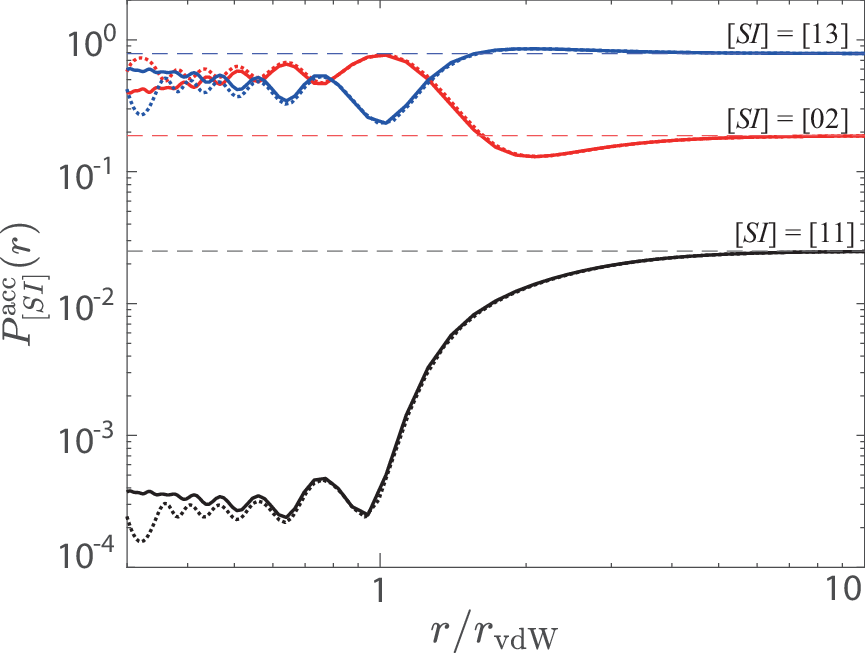} }
 \caption{\label{fig:7LiAcc} The accumulated spin fraction $P^{\rm acc}_{[SI]}$ of the scattering state for $^{7}$Li$_2$ at zero $B$-field, obtained from the original potentials (solid lines) and reduced potentials (dotted lines). The horizontal dashed lines indicate the spin fraction of 0.1875, 0.0250, and 0.7875 predicted by the basis transformation coefficient. }
\end{figure}
The spin structure of low-energy scattering and weakly bound vdW molecular states are also expected to respond to an external field, allowing for their manipulation. Given the various spin selection rules and propensity laws that govern molecular reactions, this manipulation offers promising potential for controlling these reactions. We investigate in the following the spin structure of the scattering $|\Psi_{\rm scat}\rangle$ and molecular $|\Psi_{\rm m}\rangle$ states of alkali dimers in the presence of a magnetic field and examine the performance of the reduced potential in this context. We first study the spin components of the scattering state. We shall use $^{7}$Li$_2$ at $B=0$ G as an example, which is particularly relevant for our three-body analysis \cite{SHpaper}. For that analysis, we need to introduce two types of molecular spin bases $|F M_F(f_af_b)\rangle$ and $|FM_F[SI]\rangle$, representing the different ways the atomic spins can be combined. Here, $F$ and $I$ are the quantum numbers of two-atom total spin $\vec{F}=\vec{f}_a+\vec{f}_b$ and total nuclear spin $\vec{I}=\vec{i}_a+\vec{i}_b$, respectively. The projection quantum number of $\vec{F}$ is denoted as $M_F$. (Details of the definition and relationship between $|F M_F(f_af_b)\rangle$ and $|FM_F[SI]\rangle$ spin basis are given in Appendix \ref{app1}.) 

Given that both $^7$Li atoms are in the $|1,\text-1\rangle$ state, the spin state of scattering state $|\Psi_{\rm scat}\rangle$ of $^{7}$Li$_2$ is, at large distances, the $|F M_F(f_af_b)\rangle=|2\text-2(11)\rangle$. 
In contrast, by analyzing the spin structure of the $^7$Li$_2$ molecular states $|\Psi_{\rm m}\rangle$ we have found that most of them can not be represented by a pure $|F M_F(f_af_b)\rangle$ state, unlike the case for $^{85}$Rb and $^{87}$Rb atoms, for instance. They are instead well characterized by a single $|FM_F[SI]\rangle$ state. We note that most of the molecular states of $^{23}$Na$_2$, $^{39}$K$_2$ and $^{41}$K$_2$ can also be well characterized by a single $|FM_F[SI]\rangle$ state, while the spin characterization will be more complicated for $^{133}$Cs$_2$ (see next section for more details).
Therefore, it is valuable to analyze the relationship between the spins of the scattering states and molecular states for $^7$Li as a representative example for similar species such as $^{23}$Na, $^{39}$K and $^{41}$K. This relation can play a major role in reaction processes like three-body recombination \cite{SHpaper,Haze:2022}.
As is well known, a $|F M_F(f_af_b)\rangle$ can be expressed as a linear combination of $|F M_F[SI]\rangle$ states of the same $F$ and $M_F$. For convenience, this basis transformation relation is presented in Appendix \ref{app1}. As a result, for $^7$Li$_2$, the $|FM_F[SI]\rangle$ components $P_{[SI]}$ (explicit definition for $P_{[SI]}$ is given below) of the scattering state in the whole spatial space are $P_{[02]}:P_{[11]}:P_{[13]}=0.1875:0.0250:0.7875$. However, as atoms approach interatomic distances $r \sim r_{\rm vdW}$, the spin structure of the scattering state will change. In fact, a more precise characterization of the spin structure is needed for understanding chemical reactions that typically occur within a finite interparticle volume \cite{SHpaper,Haze:2022}. Therefore, we define an accumulated spin component $P^{\rm acc}_{\eta}$  
\begin{equation}
P^{\rm acc}_{\eta}(r)\equiv \frac{\int_{0}^{r}|\langle FM_F\eta|\Psi(\vec{r}')\rangle|^2d^3\vec{r}'}{ \sum_{\eta'}\int_{0}^{r}|\langle FM_F\eta'|\Psi(\vec{r}')\rangle|^2d^3\vec{r}'},\label{Px}
\end{equation}
to quantify the spin components of the two-atom state $|\Psi\rangle$ [given by Eq.~(\ref{wf})] within a finite $r$. Here $|\Psi\rangle$ can be $|\Psi_{\rm scat}\rangle$ or $|\Psi_{\rm m}\rangle$ and $\eta=[SI]$ or $(f_af_b)$. Consequently, the full component $P_{\eta}$ is taken at the limit $P_{\eta}\equiv P^{\rm acc}_{\eta}(r\rightarrow \infty)$. Figure \ref{fig:7LiAcc} shows that the $P^{\rm acc}_{[SI]}$ calculated from our reduced potentials is in excellent agreement with that calculated from the BO potentials, in particular, when $r> 0.6$ $r_{\rm vdW}$. In the asymptotic region $r\gg r_{\rm vdW}$, both calculations reproduce the ratio $P_{[02]}^{\rm acc}:P_{[11]}^{\rm acc}:P_{[13]}^{\rm acc}=0.1875:0.0250:0.7875$ predicted by the basis transformation. However, with the decrease of $r$, the $P_{[SI]}^{\rm acc}$ starts to deviate from its asymptotic value and oscillates. Except for an opposite oscillation phase, the values of $P_{[02]}^{\rm acc}$ and $P_{[13]}^{\rm acc}$ are rather comparable at $r< 2$ $r_{\rm vdW}$, both significantly larger than that of $P_{[11]}^{\rm acc}$. This comparison has also been extended to other species, consistently showing similar agreement between results from the reduced potentials and those from the BO potentials. Nevertheless, the $r$-dependent behavior of $P^{\rm acc}[SI]$ can vary across species, particularly when $r\lesssim r_{\rm vdW}$.

 \begin{figure}[t]
 \centering
  \resizebox{0.48\textwidth}{!}{\includegraphics{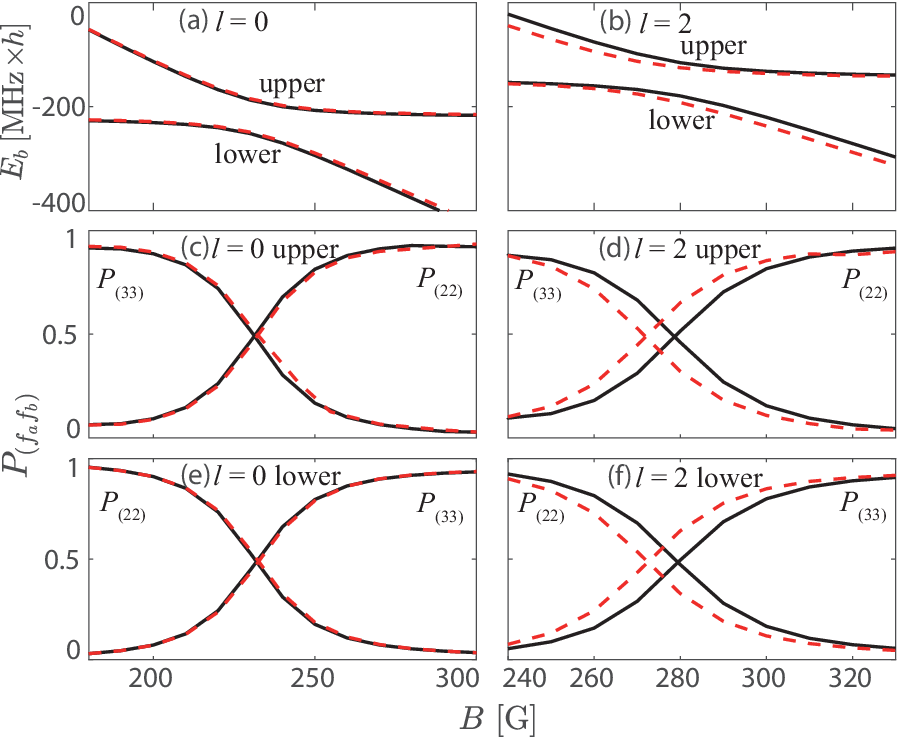} }
 \caption{\label{fig:AC} The energy of the broad $s$-wave (a) and $d$-wave (b) avoided crossing levels of $^{85}$Rb$_2$. We note that (a) and (b) are zoomed-in views of Figs. \ref{fig:85Rb2}(a) and \ref{fig:85Rb2}(b), respectively, with irrelevant levels omitted. (c) and (e) show the $|FM_F(f_af_b)\rangle$ component, $P_{(f_a f_b)}$, of the upper and lower levels in (a), respectively. (d) and (f) show $P_{(f_a f_b)}$ of the upper and lower levels in (b), respectively. The results are calculated by using the original potentials (solid lines) and reduced potentials (dashed lines).}
\end{figure}

To provide further evidence that the reduced model does properly reproduce the spin structure of the real system we now analyze the case where two molecular levels are strongly coupled to each other, for instance, in the vicinity of an avoiding crossing. This scenario is shown in Figs. \ref{fig:85Rb2}(a) and \ref{fig:85Rb2}(b) for $^{85}$Rb molecular states with $l=0$ and 2, respectively. 
In both partial waves the broad avoided crossings in Fig. \ref{fig:85Rb2} are generated by one level in $|FM_F(f_af_b)\rangle=|4\text-4(22)\rangle$ state and the other in $|4\text-4(33)\rangle$ state. We calculated the $|FM_F(f_af_b)\rangle$ spin component $P_{(f_af_b)}$ from the corresponding numerical molecular wavefunctions, $\Psi_{\rm m}(\vec{r})$. In the vicinity of the avoid crossings for both cases, the molecular levels switch their spin state between the $|4\text-4(22)\rangle$ and $|4\text-4(33)\rangle$ state, as is shown in Fig. \ref{fig:AC}. The reduced potentials reproduce the spin structure of the relevant molecular very well throughout the whole avoided crossing, in particular for the $s$-wave states. For the $d$-wave case, the reduced potentials lead to an overall shift of about 7 G in the curves of the level energy and the spin component. 

The numerical comparison so far has evidenced that our reduced potentials are a generally good replacement of the original potentials in a wide range of magnetic field regimes, despite the notable differences between the two potentials at short range. Remarkably, the reduced potentials sufficiently reproduce a variety of physical phenomena, such as the Feshbach resonance, the molecular levels avoided crossing, and the spin structure in scattering and bound molecular states. This indicates the fundamental physics governing the molecular spin structure is robust to the specific details of the short-range interactions. As we will discuss further, this understanding allows us to parameterize the complex short-range electronic spin exchange interaction, facilitating the characterization of the molecular spin structure. For deep molecular levels and those with large $l$, deviations appear on a quantitative level but leave the underlying physics qualitatively unchanged. Nevertheless, such deviations can be mitigated by considering reduced potentials supporting a larger number of molecular states. 

\section{characterizing the molecular spin structure}\label{SpinMix}
\begin{table*}
  \tabcolsep=4pt
\small
\renewcommand\arraystretch{1.0}
\caption{\label{tab:spinmixing} Interaction and spin mixing parameters for the vdW molecules of two alkali metal atoms. See Table~\ref{tab:vdWM} for the corresponding values of $r_{\rm vdW}$ and $E_{\rm vdW}$.}
 \begin{tabular}{ccccccccccc}
 \hline
 \hline
 Specie&$u_{st}$&$E_{\rm{hf}}/h$ [MHz]&$ E_{\rm{hf}}/E_{\rm{vdW}}$&$\xi_{\rm ex}$&$\phi^{\rm avg}_{(f_af_b)}$&$\phi^{\rm avg}_{[SI]}$&$\gamma_{\rm ex}$ \\ 
 \hline
 $^{7}$Li&0.3768&803.50&1.65&9.020&0.5685&0.9998&2336\\
 $^{23}$Na&0.3096&1771.63&22.81&0.6192&0.5711&0.9909&47.11 \\
 $^{39}$K&0.3984&461.72&20.81&0.7968&0.5662&0.9919&53.45 \\
 $^{41}$K&0.1231&254.01&12.34&0.3940&0.6093&0.9867&29.45 \\
 $^{85}$Rb&0.0626&3035.73& 481.86&0.1252&0.9728&0.7112&0.0942 \\
 $^{87}$Rb&0.0338&6834.68&1124.13&0.0676&0.9836&0.6232&0.0435 \\
 $^{133}$Cs&0.1387&9192.63&3455.86&0.2774&0.7937&0.6780&0.6408\\
 \hline
 \hline
 \end{tabular}
\begin{tabular}{ccccccccc}
\hline
\hline
 \end{tabular}
 \end{table*}
In this last part of our study, we aim at characterizing the molecular spin states and reveal the underlying physics controlling their composition. In fact,
the spin state of the weakly bound vdW molecules of two alkali atoms can be very different across the various atomic species. For instance, as we  discussed in the previous section, weakly bound $^7$Li$_2$ vdW molecules are rather pure in $|FM_F(SI)\rangle$ states while those for $^{85}$Rb$_2$ molecules are well characterized as $|FM_F(f_af_b)\rangle$ states. In particular circumstances, however, two $^{85}$Rb$_2$ molecular levels can also change into mixed states in the $|FM_F(f_af_b)\rangle$ basis, as highlighted in the discussions of Figs.~\ref{fig:85Rb2} and \ref{fig:AC} near an avoided crossing. In the following we ignore such particular circumstances but rather focus on the general spin structure
for a given alkali species.

To more globally characterize the molecular spin structure for a given atomic species, we define the dimensionless spin mixing parameter 
\begin{equation}
\gamma_{\rm ex} \equiv (1-\phi^{\rm avg}_{(f_af_b)})/(1-\phi^{\rm avg}_{[SI]}).\label{gex} 
\end{equation}
Here, $\phi^{\rm avg}_{(f_af_b)}$ and $\phi^{\rm avg}_{[SI]}$ denote the averaged "spin purity" in the $|FM_F(f_af_b)\rangle$ and $|FM_F[SI]]\rangle$ bases, respectively, given by
\begin{equation}
\phi^{\rm avg}_{\eta}=\frac{1}{N}\sum_{n}^N{\rm max}_{\eta \in D}\left\{\int_0^{\infty}|\langle FM_F\eta|\Psi_{\rm m}^n(r)\rangle|^2d^3\vec{r}\right\}, \label{phiave}
\end{equation}
where $\eta=(f_af_b)$ or $[SI]$ and $D$ is the corresponding variable domain of $\eta$. 
In Eq.~(\ref{phiave}) we extract the dominant spin component value of each molecular state $n$ in the $|FM_F\eta\rangle$ basis and average them over the total number $N$ of molecular states included in the reduced potentials. (A similar procedure could be performed on the molecular states of the BO potentials by restricting the range of energies of such states.)  {In this work, we consider $N\approx$ 80-200 states across the species, counting for molecular levels of all partial waves and binding energies up to a few thousands of $E_{\rm vdW}$. According to Eq.~(\ref{gex}), $\gamma_{\rm ex}\ll1$ implies that the considered molecular states are purely $|FM_F(f_af_b)\rangle$ while for $\gamma_{\rm ex}\gg1$ the  molecular states are purely $|FM_F[SI]]\rangle$, and any number in between indicates the degree of molecular spin mixing. The values for $\gamma_{\rm ex}$ and the averaged spin purities in the two bases are listed in Table~\ref{tab:spinmixing}. We find that $\gamma_{\rm ex} \ll1$ for $^{87}$Rb and $^{85}$Rb, while the $\gamma_{\rm ex} \gg1$ for $^{7}$Li, $^{23}$Na, $^{39}$K and $^{41}$K.
We note, however, that for $^{133}$Cs $\gamma_{\rm ex} \sim 1$, indicating that the Cs$_2$ molecular states are mixed in both $|FM_F(f_af_b)\rangle$ and $|FM_F[SI]]\rangle$ bases.

At zero magnetic field, the spin structure of alkali diatomic vdW molecules fundamentally originates from the competition between the electronic spin exchange and hyperfine interactions. In general, in cases where electronic spin exchange interaction is dominant over hyperfine interaction molecular states are well characterized as a $|FM_F[SI]]\rangle$ state. On the other hand, for cases where the hyperfine interaction is dominant over electronic spin exchange interaction molecular states are instead of $|FM_F(f_af_b)\rangle$ character. Evidently, in the case where there is a balance between these two interactions, the molecular states can be mixed in both $|FM_F[SI]]\rangle$ and $|FM_F(f_af_b)\rangle$ basis. In the following, we aim to gain a deeper understanding of our numerical results on the spin mixing parameter $\gamma_{\rm ex}$ of vdW molecules by characterizing both electronic exchange and hyperfine interactions. We shall focus on a simple, and physically intuitive, picture for qualitative understanding. 

\subsection{Effective electronic spin exchange interaction}
The electronic spin exchange interaction 
corresponds to the difference between singlet and triplet potentials which depends on $r$, which is prominent at short-range. Its actual form can be complex and can vary significantly across atomic species. Nevertheless, our comparative analysis indicates that the molecular spin structure is insensitive to the specific details of the interatomic interaction at short range. Largely different potential models that share the same singlet and triplet scattering lengths or bound levels provide comparable descriptions of the spin structure in weakly bound molecules. 
This motivates us to effectively parameterize the complex short-range electronic spin exchange interaction using the energy splittings between singlet and triplet levels, which result from the electronic spin exchange interaction. Specifically, we will define an effective electronic spin exchange interaction for a given vdW molecular state as the energy difference between adjacent singlet and triplet levels.
We focus on $s$-wave vdW molecular states.

For vdW interactions, molecular levels can be classified according to their location within a given universal energy range for each vibrational quantum number $v$ of a given partial wave $l$ \cite{chin:2010, Gao:2000}. For instance, for $l=0$, the $v=-1$ level is always within a energy range of [-39.5, 0] $E_{\rm vdW}$ while the $v=-2$ level is within [-272.5, -39.5] $E_{\rm vdW}$. These universal energy intervals are referred to as "vdW energy bins" \cite{chin:2010} and are illustrated in Fig.~\ref{fig:EST} by the different colored regions. According to quantum defect theory \cite{Gao:2000}, for vdW interactions the location of the energy of an $s$-wave level in a given energy bin can be characterized in terms of the physical scattering lengths ($a_s$ or $a_t$), through the quantity $u$ defined as \cite{Friedrich:2004}    \begin{equation}
u(a)=\tan^{-1}[\bar{a}/(a-\bar{a})]/\pi, \label{ua}
\end{equation}
where $a$ denotes the scattering length of the associated interaction potentials and $\bar{a}\approx0.9560$ $r_{\rm vdW}$, the mean scattering length \cite{Flambaum:1999}. For $a \rightarrow \pm \infty$, $u$ approaches zero and the ($v=-1$) bound level approaches the potential threshold (i.e., the upper boundary of the first bin) while the energy of additional more deeply bound molecular states approach the other boundaries of the energy bins. For finite values of $a$, $u$ is non-zero and characterizes the distance of the bound levels from such a boundary. Figure~ \ref{fig:bin} illustrates the behavior of the bound levels in the first three energy bins with a varying $a$ tuned by the short-range parameter $\lambda$ for the Lennard-Jones potential $V_{LJ}(r)=-C_6/r^6(1-\lambda^6/r^6)$. This figure clearly shows that the bound levels appear at the boundaries of the bins when $a \rightarrow \pm \infty$, while moving away from the boundaries when $a$ is finite.

Since $u$ represents the relative position of a bound level within a specific energy bin, the energy separation between two bound levels can be approximated by multiplying the difference in 
$u$ by the bin size. For nearby singlet and triplet levels, their energy separation, and consequently the effective electronic spin exchange interaction, can be characterized by $|u(a_s)-u(a_t)|\in [0,1]$. We note that $|u(a_s)-u(a_t)|$ will properly parameterize the energy difference between nearby singlet and triplet levels whenever they are in the same energy bin or adjacent bins. A concrete example of the latter is $^{85}$Rb$_2$ in Fig. \ref{fig:EST}, in which $|u(a_s)-u(a_t)|=0.06$ characterized properly the small energy difference between a singlet level at the top of one bin and a triplet level at the bottom of the adjacent bin from above.
However, the energy separation of nearby singlet and triplet levels are not fairly described by $|u(a_s)-u(a_t)|$ when the values for $a_s$ and $a_t$ are close to each other but with one slightly smaller and the other slightly larger than $\bar{a}$. 
This problem can be resolved by defining 
$u_{st}=\min(|u(a_s)-u(a_t)|, 1-|u(a_s)-u(a_t)|)\in [0,1/2]$. 
The values of $u_{st}$ for different atomic species we study here are listed in Table \ref{tab:spinmixing}. These values are generally consistent with the energy separation of singlet and triplet levels in Fig. \ref{fig:EST}. As a result, for the $i$th energy bin, the quantity 
\begin{align}
\tilde{E}_{\rm ex}^{(i)}=u_{st}E_{\rm bin}^{(i)},
\end{align}
will provide a measure of the effective electronic spin-exchange interaction, as a simple alternative to the actual energy difference between nearby singlet and triplet levels, $\Delta E_{st}^{(i)}$, calculated numerically for a given potential model. In Fig. \ref{fig:ust}, we use the energies from the BO potentials to demonstrate that $\tilde{E}_{\rm ex}^{(i)}$ is a fairly good approximation to $\Delta E_{st}^{(i)}$. This validates the use of $\tilde{E}_{\rm ex}^{(i)}$ to characterize the effective exchange interaction of a given energy bin. 

 \begin{figure}[t]
 \centering
  \resizebox{0.5\textwidth}{!}{\includegraphics{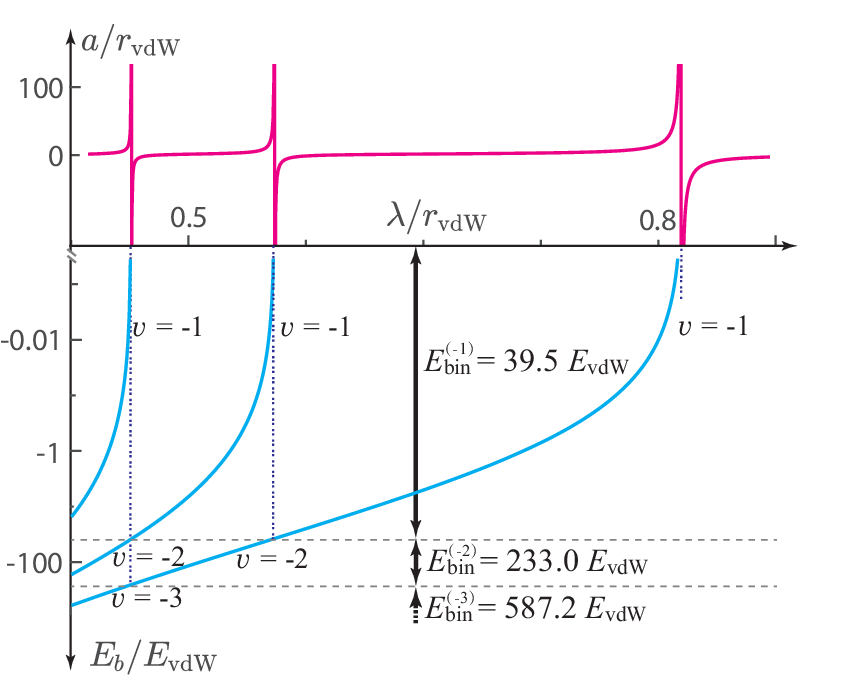} }
 \caption{\label{fig:bin} The scattering length and bound-state energies of a Lennard-Jones potential are tuned by the short-range parameter $\lambda$. The horizontal solid lines represent the boundaries of vdW energy bins. The dotted vertical lines indicate that vdW bound levels approach these boundaries when the two-body scattering length diverges.}
\end{figure}

 \begin{figure}[t]
 \centering
  \resizebox{0.48\textwidth}{!}{\includegraphics{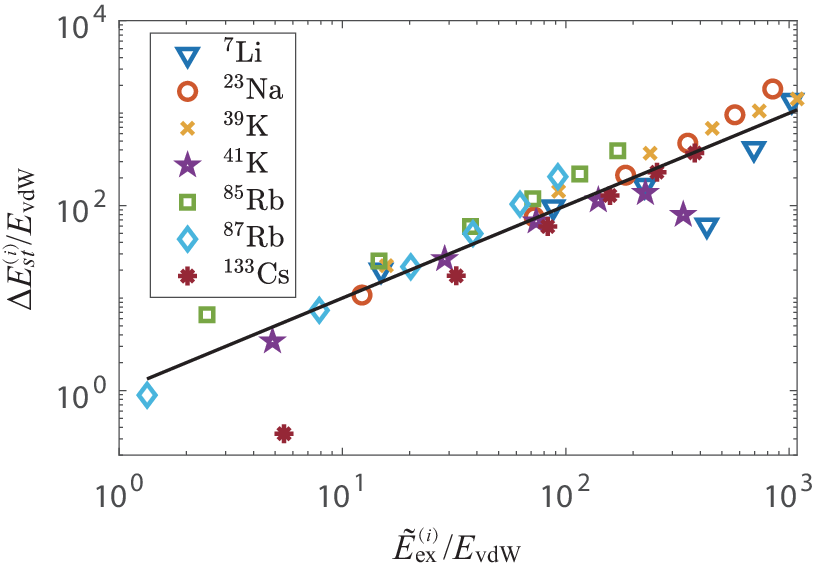} }
 \caption{\label{fig:ust} The energy difference of the nearby singlet and triplet levels $\Delta E_{st}^{(i)}$versus $\tilde{E}_{\rm ex}^{(i)}$ in the first six energy bins. The solid line marks where $\Delta E_{st}^{(i)}=\tilde{E}_{\rm ex}^{(i)}$.}
\end{figure}

\subsection{Effective hyperfine interaction}

In order to characterize the effective hyperfine interaction we follow a similar approach to that of the electronic spin exchange interaction. Now we focus on the energy difference between nearby molecular states in the presence of the hyperfine interaction.
We consider a given molecular level in the ground hyperfine state $|FM_F(f_a f_b)\rangle$ =$|2f^*,\text-2f^*(f^*f^*)\rangle$ as a reference. The energy of the corresponding molecular levels of the excited hyperfine state will be shifted upwards by, approximately, the hyperfine splitting constant $E_{\rm hf}=2(f^*+1)A_{\rm hf}$ if one atom is in the excited $f^*+1$ state, or $2E_{\rm hf}$ in the case that both atoms are in the $f^*+1$ state. Based on this, one could intuitively characterize effective hyperfine interaction from the energy difference of nearby hyperfine molecular levels, $E_{\rm hf}$. This, however, is valid only if $E_{\rm hf}$ is small comparable to the energy bin so that the original and the shifted energy levels are still in the same energy bin.
Nevertheless, in some cases the hyperfine interaction can shift a level into a different energy bin if $E_{\rm hf}$ is large enough
and the energy difference of nearby hyperfine molecular levels is not simply $E_{\rm hf}$ or $2E_{\rm hf}$. Instead, it is restricted to the size of the corresponding energy bin with an upper bound given, roughly, by $E_{\rm bin}/2$. Therefore, as a more precise characterization of the energy separation is not straightforwardly available, we will utilize this upper bound for a rough characterization of the effect of the hyperfine interaction. Taking both cases into account, we use the quantity
\begin{align}
    \tilde{E}_{\rm hf}^{(i)}= \min(E_{\rm hf}, E_{\rm{bin}}^{(i)}/2),
\end{align} 
to characterize the effective hyperfine interaction in the $i$th energy bin. The values for  
$E_{\rm hf}$ for different atomic species we study here are listed in Table \ref{tab:spinmixing}.

\subsection{The exchange parameter}
According to the above analysis, we can now define the dimensionless exchange parameter as a ratio between effective electronic spin exchange and effective hyperfine interactions
\begin{equation}
\xi_{\rm ex}^{(i)}=\frac{\tilde{E}_{\rm ex}^{(i)}}{\tilde{E}_{\rm hf}^{(i)}}=\frac{u_{st}E_{\rm bin}^{(i)}}{\min(E_{\rm hf}, E_{\rm{bin}}^{(i)}/2)}
\end{equation}
to estimate the relative strength of the electronic spin exchange for vdW molecules. This definition for $\xi_{\rm ex}^{(i)}$ depends on the energy bin $i$ in which the molecular states are located when used to characterize the spin structure of the corresponding molecular states. Figure \ref{fig:Xi} shows that the value of $\xi_{\rm ex}^{(i)}$ increases with the bin index $i$ for $^7$Li, $^{23}$Na, $^{39}$K and $^{41}$K while for heavier species $^{85}$Rb, $^{87}$Rb and $^{133}$Cs such bin-dependency is rather weak, particularly in the first few bins relevant to our present study. Nevertheless, the values of $\xi_{\rm ex}^{(i)}$s across different species preserve the same order from the first bin down to the 5th bin, with relative values consistent to those of $\gamma_{\rm ex}$. For deeper bins, the order switches only between $^{85}$Rb and $^{133}$Cs. This demonstrates the qualitative agreement between our simple parameter $\xi_{\rm ex}^{(i)}$ and the numerical results $\gamma_{\rm ex}$ to characterize the spin structure of vdW molecules. For instance, in the cases of strong spin exchange ($\xi_{\rm ex}^{(i)}\gg 1$), as is the case of $^7$Li, we also obtain $\gamma_{\rm ex}\gg1$, for weak spin exchange species $^{85}$Rb and $^{87}$Rb, we find both $\xi_{\rm ex}^{(i)}\ll 1$ and $\gamma_{\rm ex}\ll1$, while for intermediate spin exchange species, like  $^{133}$Cs, we also have $\xi_{\rm ex}\sim 1$ and $\gamma_{\rm ex}\sim1$. It should be noted that using $ \tilde{E}_{\rm hf}$ instead of $E_{\rm hf}$ in the definition of $\xi_{\rm ex}^{(i)}$ is crucial for explaining the $\gamma_{\rm ex}$ for $^{133}$Cs. For $^{23}$Na, $^{39}$K and $^{41}$K atoms, the corresponding vdW molecules evolve from the intermediate to a strong spin exchange regime with the increase of bin index $i$. We also emphasize here the advantage that $\xi_{\rm ex}^{(i)}$ can reveal the physical origin of the molecular spin property based on fundamental atomic and interatomic interactions. We note that, alternatively, the spin structure of vdW molecules can be analyzed using the widely recognized framework of Hund's cases. In this context, an intriguing question arises about how to relate our parameter $\xi_{\rm ex}^{(i)}$ to Hund's cases, a topic that merits further investigation. In Table \ref{tab:spinmixing}, we list the minimal value of $\xi_{\rm ex}^{(i)}$ (i.e. that of the first bin) with the superscript omitted. This value is used in the analysis presented in Ref. \cite{SHpaper}.

\begin{figure}[htbp]
	\includegraphics[width=0.95\columnwidth]{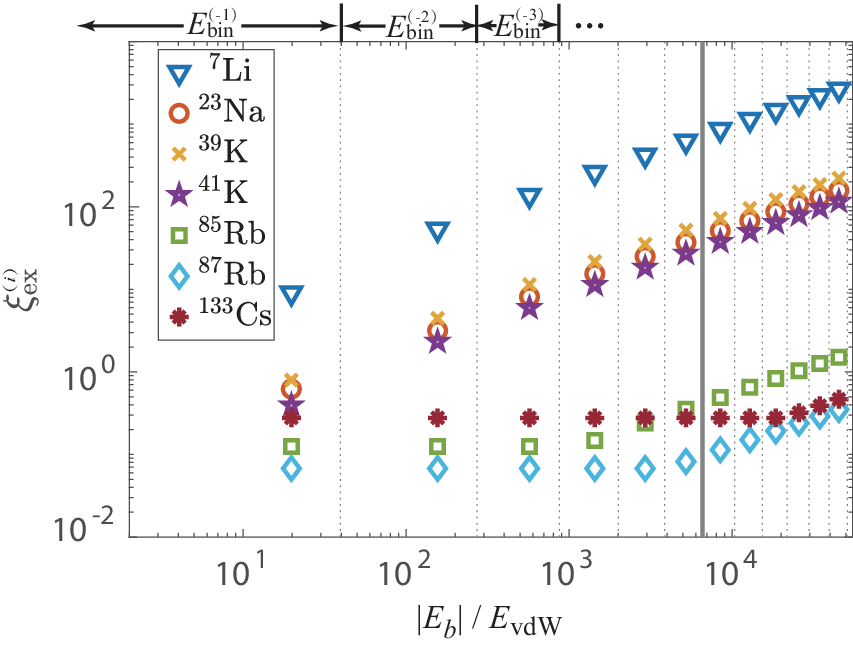}
	\caption{\label{fig:Xi} $\xi_{\rm ex}$ versus vdW energy  bins. The data points of each bin are placed at the corresponding middle value of $|E_{b}|$. The vertical lines indicate the boundaries of the bins. The right boundary of the sixth energy bin is highlighted as it is the binding energy regime covered by our reduced potential models (except for that for $^{133}$Cs).}
\end{figure}

\section{conclusion and outlook}

In summary, we have studied the weakly bound van der Waals molecules of $^7$Li$_2$, $^{23}$Na$_2$, $^{39}$K$_2$, $^{41}$K$_2$, $^{85}$Rb$_2$, $^{87}$Rb$_2$ and $^{133}$Cs$_2$, with a focus on determining and understanding their binding energies and spin structures. For each atomic species, we constructed reduced singlet and triplet potentials to replace the original Born-Oppenheimer potentials. The reduced potentials support typically only $5\sim 7$ $s$-wave bound states and are more suitable for numerical simulations of various three-body problems. We show that they can reproduce well the physical properties of two-atom scattering and molecular states in the presence of a magnetic field, such as the scattering length, the scattering effective range and the molecular binding energy as well as the spin structure of scattering and molecular wavefunctions. At zero magnetic field, we define the quantity $\gamma_{\rm ex}$ to characterize the molecular spin structure by using the averaged spin purity in two different bases, obtained directly from the numerical simulations for each atomic species. The result of $\gamma_{\rm ex}$ across alkali species characterizes the competition between the electronic spin exchange and the hyperfine interactions. For van der Waals molecules, we define a simple parameter $\xi_{\rm ex}$, which is a function of the singlet and triplet scattering lengths, the atomic hyperfine splitting constant, and the size of the energy bin for a given molecular level. We find that  $\xi_{\rm ex}$ captures qualitatively the competition between the effective electronic spin exchange and hyperfine interactions and explains fairly well the numerical result of $\gamma_{\rm ex}$.

For future studies, the characterization of spin structure of vdW molecules can be extended to a finite magnetic field to include the Zeeman interaction. The understanding of vdW molecules that we have gained here can be applied to explore the underlying mechanisms of reactions that involve these molecules. The result of the present work has been already applied in our studies on the spin hierarchy in the product state distribution of three-body recombination \cite{SHpaper} and on controlling three-body recombination reaction via a Feshbach resonance \cite{FRpaper}. Further studies on the control of the three-body recombination via the spin mixing of product molecular state \cite{Dorer:2025} as well as on ultracold state-to-state atom-vdW molecule reactions are underway. 

This work was supported by the Baden-W\"urttemberg Stiftung through the Internationale Spitzenforschung program 
(BWST, contract No.~ISF2017-061) and by the German Research Foundation (DFG, Deutsche Forschungsgemeinschaft, 
contract No. 399903135). We acknowledge support from bwForCluster JUSTUS 2 for high performance computing. J.H.D and J.P.D. acknowledge funding by Q-DYNAMO (EU HORIZON-MSCA-2022- SE-01) within project No. 101131418. 
J.P.D. also acknowledges partial support from the U.S. National Science Foundation (PHY-2012125 and PHY-2308791) 
and NASA/JPL (1502690).

\newpage

\appendix
\section{Molecular spin basis} \label{app1}
Following standard quantum mechanics, the two molecular bases $|FM_F(f_af_b)\rangle$ and $|FM_F[SI]\rangle$ can be expressed as
\begin{align}
    |FM_F(f_af_b)\rangle=\sum_{m_{f_a}m_{f_b}}C_{f_am_{f_a}f_bm_{f_b}}^{FM_F}|f_am_{f_a}\rangle|f_bm_{f_b}\rangle,
\end{align} 
and  
\begin{align}
    |FM_F[SI]\rangle=\sum_{M_{I}M_{S}}C_{SM_SIM_I}^{FM_F}|SM_S\rangle|IM_I\rangle,
\end{align}
respectively, where $C_{j_1m_1j_2m_2}^{j_3m_3}$ is the Clebsch-Gordan coefficient. We use $M_S$ and $M_I$ to denote the projection quantum number of the molecular total electronic and nuclear spins, respectively. 

The $|FM_F(f_af_b)\rangle$ basis is connected to the $|FM_F[SI]\rangle$ basis via a $9\text-j$ symbol
\begin{eqnarray}
 |FM_F(f_a f_b)\rangle=\sum_{S,I}&\sqrt{2 f_a+1}\sqrt{2 f_b+1}\sqrt{2S+1}\sqrt{2I+1} \notag \\
&\times \sqrt{2-\delta_{f_a f_b}} \left\{ 
\begin{array}{ccc}
     s_a&s_b&S  \\
     i_a&i_b&I  \\
     f_a&f_b&F 
\end{array}
\right\}
|FM_F[SI]\rangle.\label{eq:9j} \notag \\
\end{eqnarray}
Applying Eq. (\ref{eq:9j}) to our investigation of $^7$Li gives the expression of the initial state of the reacting atomic pair
\begin{equation}
|2\text-2(11)\rangle=\frac{\sqrt{3}}{4}|2\text-2[02]]\rangle-\frac{1}{2\sqrt{10}}|2\text-2[11]\rangle+\frac{3\sqrt{7}}{4\sqrt{5}}|2\text-2[13]\rangle,
\end{equation}
which leads to the ratio of the initial components in $|2\text-2[02]\rangle$, $|2\text-2[11]\rangle$ and $|2\text-2[13]\rangle$
\begin{equation}
\left|\frac{\sqrt{3}}{4}\right|^2:\left|-\frac{1}{2\sqrt{10}}\right|^2:\left|\frac{3\sqrt{7}}{4\sqrt{5}}\right|^2=0.1875:0.0250:0.7875,
\end{equation}
respectively. 
\newpage

\begin{thebibliography}{59}
\bibitem [{3bn()}]{3bnum}%
  \BibitemOpen
  \href@noop {} {}\bibinfo {note} {Three-body simulation is not only essential
  for three-body recombination but also necessary for the reaction of a vdW
  molecule and an atom because the vdW bond is very loose.}\BibitemShut {Stop}%
\end{thebibliography}
%

\end{document}